\documentclass[fleqn,usenatbib]{mnras}

\usepackage{mathptmx}

\usepackage[T1]{fontenc}
\DeclareUnicodeCharacter{00A0}{~}

\DeclareRobustCommand{\VAN}[3]{#2}
\let\VANthebibliography\thebibliography
\def\thebibliography{\DeclareRobustCommand{\VAN}[3]{##3}\VANthebibliography}

\usepackage{graphicx}	
\usepackage{amsmath}	
\usepackage{amssymb}	
\usepackage{float}

\usepackage{multirow}

\usepackage{longtable}
\usepackage{rotating}
\graphicspath{ {./plots/} }
\usepackage{geometry}
\usepackage{pdflscape}
\usepackage{hyperref}
\usepackage{tablefootnote}
\usepackage{soul}
\usepackage{threeparttable}
\usepackage{multicol,lipsum}
\usepackage{lineno}

\title[\textit{TESS} Blazars]{Exploring Short-Term Optical Variability of Blazars Using \textit{TESS}}

\author[Pininti et al.]{Vivek Reddy Pininti$^{1,2}$,
Gopal Bhatta$^{3}$,
Sagarika Paul$^{4}$,
Aman Kumar$^{5}$,
Aayushi Rajgor$^{6}$,
Rahul Barnwal$^{7}$,
\newauthor and Sarvesh Gharat$^{8}$
\\
$^{1}$Universit\"{a}t Potsdam, Institut f\"{u}r Physik und Astronomie, Karl-Liebknecht-Stra{\ss}e 24/25, 14476 Potsdam, Germany\\
$^{2}$Leibniz-Institut f\"{u}r Astrophysik Potsdam (AIP), An der Sternwarte 16, 14482 Potsdam, Germany\\
$^{3}$Institute of Nuclear Physics, Polish Academy of Sciences, PL-31342 Krak\'{o}w, Poland\\
$^{4}$Indian Institute of Science Education and Research (IISER) Mohali, Punjab 140306, India\\
$^{5}$Tezpur University, Tezpur, Assam 784028, India.\\
$^{6}$K. J. Somaiya College of Engineering, Mumbai, Maharashtra 400077, India\\
$^{7}$Indian Institute of Information Technology, Kottayam, Kerala 686635, India\\
$^{8}$C-MInDS, Indian Institute of Technology Bombay, Powai, Maharashtra 400076, India
}

\date{Accepted XXX. Received YYY; in original form ZZZ}

\pubyear{2021}

\begin{document}
\label{firstpage}
\pagerange{\pageref{firstpage}--\pageref{lastpage}}
\maketitle

\begin{abstract}
We present a first systematic time series study of a sample of blazars observed by the Transiting Exoplanet Survey Satellite \textit{TESS} spacecraft. By cross matching the positions of the sources in the TESS observations with those from Roma-BZCAT, 29 blazars including both BL Lacerate objects and flat-spectrum radio quasars were identified. The observation lengths of the 79 light curves of the sources, across all sectors on which the targets of interest have been observed by \textit{TESS}, range between 21.25 and 28.2 days. The light curves were analyzed using various methods of time series analysis. The results show that the sources exhibit significant variability with fractional variability spanning between 1.41\% and 53.84\%. The blazar flux distributions were studied by applying normal and lognormal probability density function models. The results indicate that optical flux histogram of the sources are consistent with normal probability density function with most of them following bi-modal distribution as opposed to uni-modal distribution. This suggests that the days-timescale optical variability is contributed either by two different emission zones or two distinct states of short-term activity in blazars. Power spectral density analysis was performed by using the power spectral response method and the true power spectra of unevenly sampled light curves were estimated. The power spectral slopes of the light curves ranged from 1.7 to 3.2.

\end{abstract}




\section{Introduction}


Imaging of most galaxies shows that their observed optical luminosities are due to the constituent stars that are almost uniformly distributed throughout the galaxy. Such galaxies are called normal galaxies. On the other hand, a fraction of all observed galaxies emit substantial radiation from their central region or nuclei and thereby outshine the entire galaxy. The core region of such active galaxies are widely known as Active Galactic Nuclei (AGN). These sources are one of the most luminous objects in the universe with their luminosities of the order of $10^{47}$ ergs/s. Historically, the distant AGNs were mistaken with the nearby stars; hence they were termed as quasar.

At the time of their discovery around 1963 \citep{1999PASP..111..661S}, scientists quickly realized that the high luminosity of AGNs could not be powered from nuclear processes. Today, it is widely accepted that such galaxies host super-massive black holes (SMBH) at the center with gravitational potential strong enough to pull surrounding matter comprised of gas, dust and stars to form an accretion disc. The combination of strong gravitational potential and magnetic fields from these BHs and loss of angular momentum of infalling materials result in the release of tremendous amount of energy that is sufficient to support bi-polar relativistic jets. The jets transport a part of the total energy into the intergalactic medium in the form of kinetic energy of the constituent particles.

A small fraction of AGNs ($\sim10\%$) \citep{urry1995unified} extensively emit in radio frequency and are known as radio-loud sources. Blazars are a sub-population of radio-loud AGNs that expel powerful relativistic jets closely aligned to the line of sight. The term `blazar' was first introduced by Edward Spiegel at the Pittsburgh Conference on BL LAC Objects in the year 1978 \citep{pittir27771}. On the basis of equivalent width (EW) of the optical emission lines \citep{urry1995unified}, Flat Spectrum Radio Quasars ({FSRQ}) along with the {BL Lac}ertae (BL Lac) objects normally constitute the blazar class of AGNs. These two radio galaxy types are widely known to be distinct in morphology such that the parent population of FSRQs are the Fanaroff-Riley type II radio galaxies (FRIIs), and that of BL Lacs are the FRIs \citep[for some exceptions, see e. g.,][]{Kharb2010, Rector2001}.

Blazars are one of the most variable extra-galactic objects with a strong broadband emission ranging from radio to TeV energies.
The emitted non-thermal radiation exhibits rapid flux and polarization variability on diverse timescales, typically a few minutes to decades. Strong Doppler boosted relativistic jets that are viewed at small angles, i.e., $\lesssim5^{\circ}$ for BL Lac and $\lesssim10^{\circ}$ for FSRQ, cause relativistic beaming which is responsible for dominant radiation processes in blazars \citep{1999ApJ...521..493L}. Such a Doppler beaming of the jets amplifies the apparent luminosity and shortens the apparent variability time scales, lending the sources distinct properties that can be analyzed and constrained in the framework of time-domain astrophysics. As a result, the study of flux change from blazars become ideal probe to the central regions of AGN.

On the basis of the presence of emission lines over the broadband continuum emission, blazars are grouped into two subclasses. The more luminous type is know as flat-spectrum radio quasars (FSRQ), which show emission lines over the continuum. The synchrotron peak of the sources lies in the lower frequency; and the external Compton (EC) scenario is thought to be more plausible scenario in explaining the high-energy emission \citep{fossati1998unifying}. This is because FSRQs have abundant seed photons originating from Accretion Disc (AD), Broad Line Region (BLR), and Dusty Torus (DT). On the other hand, the less powerful type of sources, known as BL Lac objects, show weak or no such lines. In these sources, the peak of the synchrotron emission lies in the optical to X-ray regions, and the origin of high energy emission, ranging from tens of keV to TeV energies, is most likely due to the synchrotron-self Compton (SSC) effect. Possible reasons for the apparent low luminosities (compared to FSRQs) of these sources are the lack of strong circumnuclear photon fields and comparatively low accretion rates. Apart from FSRQ-BLLac dichotomy, blazars can also be classified according to their synchrotron peak frequency ($\nu_\mathrm{peak}^S$). According to this scheme of classification, blazars can be high synchrotron peaked blazars (HSP; $\nu_\mathrm{peak}^S > 10^{15} $ Hz), intermediate synchrotron peaked blazars (ISP; $10^{14} < \nu_\mathrm{peak}^S < 10^{15}$ Hz), or low synchrotron peaked blazars (LSP; $\nu_\mathrm{peak}^S < 10^{14}$Hz) \citep{2010ApJ...716...30A}.

Blazar variability in general is observed to be erratic in nature. But the source light curves display persistent activity along with superimposed intermittent flares distinguished by a well-defined rise and decline in the flux level. One of the defining properties of blazar is the variability over diverse time scales. The observed variability broadly can be categorized as intra-day variability (IDV) where timescales range from a few minutes to several hours and flux changes by a few tenths of magnitude (e.g. \citet*{1995ARA&A..33..163W}). Short term variability (STV) can be observed in the timescales ranging from several days to months, whereas Long Term variability (LTV) persist over the timescales ranging from a few months to decades \citep{fan2009radio,gupta2016multiband,gaur2015optical}. In addition to the aperiodic flux variability, several sources are known to show quasi-periodic modulations in their multi-wavelength (MWL) observations \cite[see][and the references therein]{Bhatta2016}.

The literature is full of several blazar models that possibly can shape the variability observed in the MWL observations of the sources. The models can be broadly classified into two groups: In the \emph{intrinsic} scenario the flux modulations in the sources can be ascribed to a number of processes e. g. the interaction of shocks with the turbulent regions or magnetic field anomalies existing in the jet \citep[e. g.][]{marscher2013turbulent,calafut2015modeling}, or the generation of ultra-relativistic mini-jets within the jet \citep[e.g][]{giannios2009fast}.
On the other hand, according to the extrinsic models of variability projection, geometrical effects could play dominant role in producing the observed variability features. For example, as the viewing angle to a moving, discrete emitting region changes, Doppler boosting of the emitted radiation also changes \citep[e. g.][]{raiteri2013awakening,Larionov2010}. Also the geometrical effects could play to result in an overall bending of the jets, either through instabilities \citep{Pollack_2016}, or through orbital motion \citep{valtonen2012optical}.
Blazar variability over longer timescales also could be a combination of both intrinsic and extrinsic mechanisms: shocks propagating down the twisted jets and plasma blobs traveling through some helical structure in the magnetized jets are some of the examples \citep[see][]{Raiteri2017,Camenzind1992}.

A detailed study of optical variability can add to the understanding of blazar processes. For example, using variability studies jet structures and location of the emitting region in relation to the central engine can be estimated. Variability at the shortest measurable timescales should provide information on emitting regions at the smallest spatial scales. This can be used to test any relation with the emission at higher frequencies. Optical variability of blazars in various timescales have been studied mainly using the ground-based telescopes: intraday timescales \citep{Webb2021,Kalita2021,galaxies6010002,Heidt1996,1995ARA&A..33..163W}, long-term timescale \citep{2021ApJ...923....7B,Valverde2020,Nilsson2018,Fan2009}

In this paper, optical variability properties of 29 blazars using high cadence and largely continuous observations from The Transiting Exoplanet Survey Satellite (\textit{TESS}) have been studied. The paper is structured as follows: In Section 2, the \textit{TESS} data acquisition and processing is described. In Section 3 we present the \textit{TESS} light curves. Section 4 the time series analysis of \textit{TESS} data and the results are presented. In Section 5, we discuss some of the possible scenarios to explain the observed results. Finally, Section 6 summarizes our work.

\section{Observations by \textit{TESS}}
Launched on April 18, 2018, the Transiting Exoplanet Survey Satellite's (\textit{TESS}) mission is to find exoplanets orbiting the brightest dwarf stars \citep{2015JATIS...1a4003R}. In addition, the primary mission allows scientists from a wider astrophysics community to request targets for research on roughly 10,000 additional objects through each cycle of its Guest Investigator program \footnote{\url{https://heasarc.gsfc.nasa.gov/docs/tess/primary.html}}.
\textit{TESS} instruments \footnote{\url{https://heasarc.gsfc.nasa.gov/docs/tess/documentation.html}} consist of four wide-field charge-coupled device (CCD) cameras that can image a region of the sky measuring $24^{\circ} \times 96^{\circ}$. For a minimum of 27 days and maximum of 356 days, the cameras estimate the brightness of $\sim$15,000-20,000 stars every two minutes. \textit{TESS} science data have a nominal cadence of 2 minutes. The Full Frame Images (FFIs) are calibrated by the Science Processing Operations Center (SPOC) and are collected at 30 minutes cadence. This data is then made available to the public in the form of Target Pixel Files (TPFs) and calibrated light curves. \textit{TESS} has its own catalog of data and it contains mostly observations taken by \textit{TESS} for luminous objects. The data collected by \textit{TESS} is available in MAST portal or through Astroquery \citep{astroquery}. There is a substantial difference in the trend between the light curves obtained from Simple-Aperture Photometry (SAP) and the Pre-search Data Conditioning SAP (PDCSAP). PDCSAP flux is filtered and processed to remove trends that are known to affect exoplanet detection. This implies the removal of long term trends by using Co-trending Basis Vectors (CBVs). However, this process, in case of blazars, may remove intrinsic source variability. To check the stability of \textit{TESS} photometry, the light curves were also extracted from the Full Frame Images (FFI) by manually selecting the aperture around a blazar source and then plotting the light curve. However, even slight difference in aperture choice led to drastically different light curves and hence the choice to use SAP flux for our study was made. Also, \citet{2021MNRAS.501.1100R} checked the reliability of the SAP flux for S5 0716+714 by comparing it with ground-based WEBT data and found that errors for SAP fluxes are small and range from 0.3 to 0.6 per cent. The sector wise details of the objects studied in this paper are listed in Table~\ref{tab:sectordates}.

\section{DATA}
\begin{figure}
\centering
\includegraphics[width=\columnwidth]{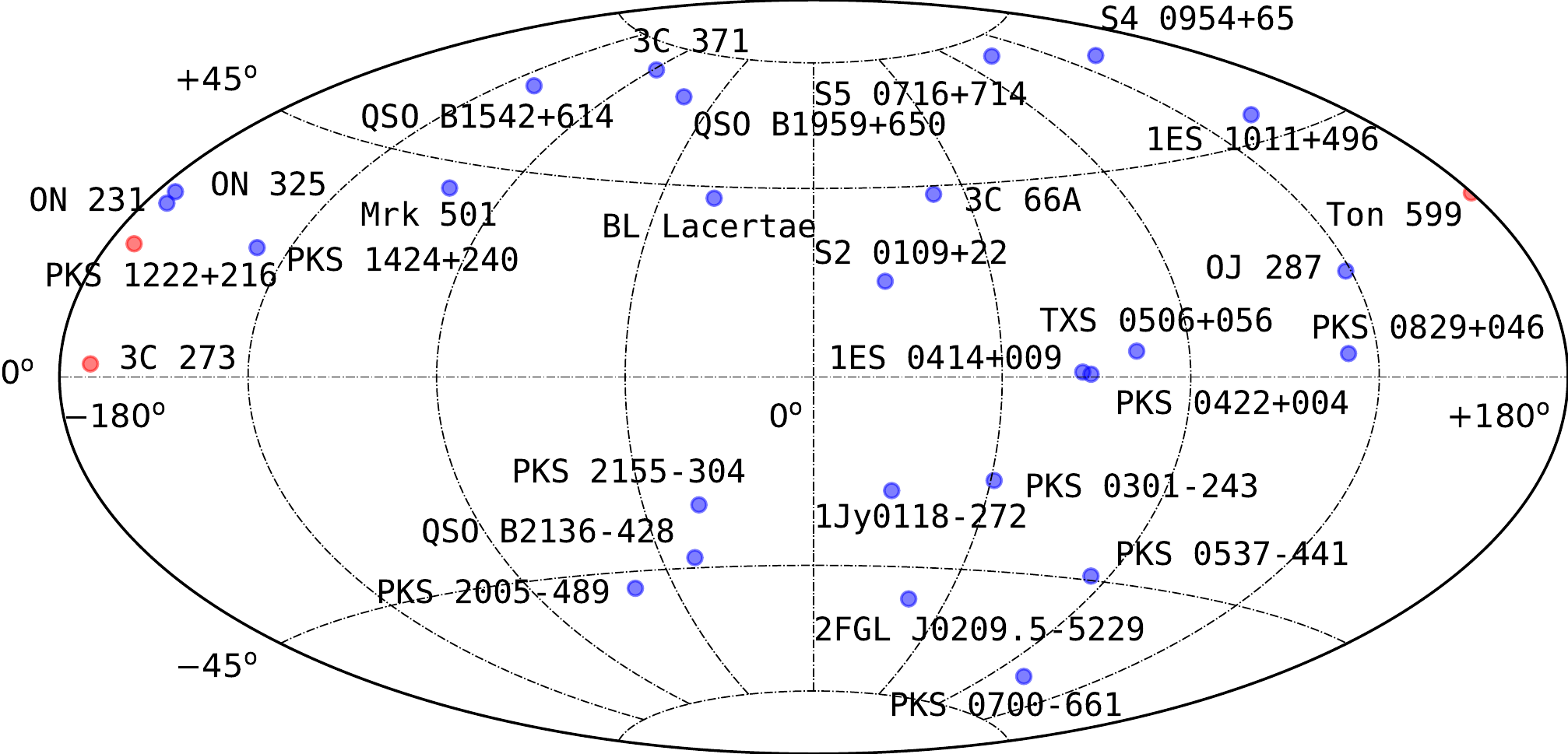}
\caption{The sky distribution of the sources that are identified as blazars by cross-matching with the Roma-BZCAT Multi-Frequency Catalogue of Blazars. The source locations are presented in the International Celestial Reference System (ICRS) coordinates with origin at Solar System Barycenter. The sub-groups of of blazars, BL Lac objects and FSRQs, are distinguished by the blue and red symbols, respectively}
\label{fig:sky-distr}
\end{figure}

We have considered the blazars from the fifth edition of the Roma-BZCAT Multi-Frequency Catalogue of Blazars \footnote{\href{https://www.asdc.asi.it/bzcat/}{The Roma BZCAT - 5th edition}}\citep{massaro2009roma} and looked at the \textit{TESS} catalog to check which of these blazars have been observed by the \textit{TESS} spacecraft. Before analyzing the light curves from \textit{TESS}, there was a need to confirm whether the sources of the light curves are of the targeted blazars. The light curve sources were validated by using the associated FITS photos to cross-reference the sources with the \textit{GAIA} catalog. The sky distribution of the cross-matched sources that are identified as blazars is presented in Figure~\ref{fig:sky-distr}.

Once it was certain that our objects of interest were within the default aperture, we have considered the SAP fluxes with 2 minute cadence to conduct our study. Besides the 2 minute exposure, SPOC pipeline also has SAP flux with 20 second exposure, starting from 2021 \citep{2021tsc2.confE.183J}. Few blazars have been observed by the \textit{TESS} multiple times i.e., they have been observed in multiple sectors. This brings the count of light curves studied here to 79, for the 29 blazars. Table~\ref{tab:sectordates} presents information about the sectors of the blazars included this study.

\begin{figure}
\centering
\includegraphics[width=\columnwidth]{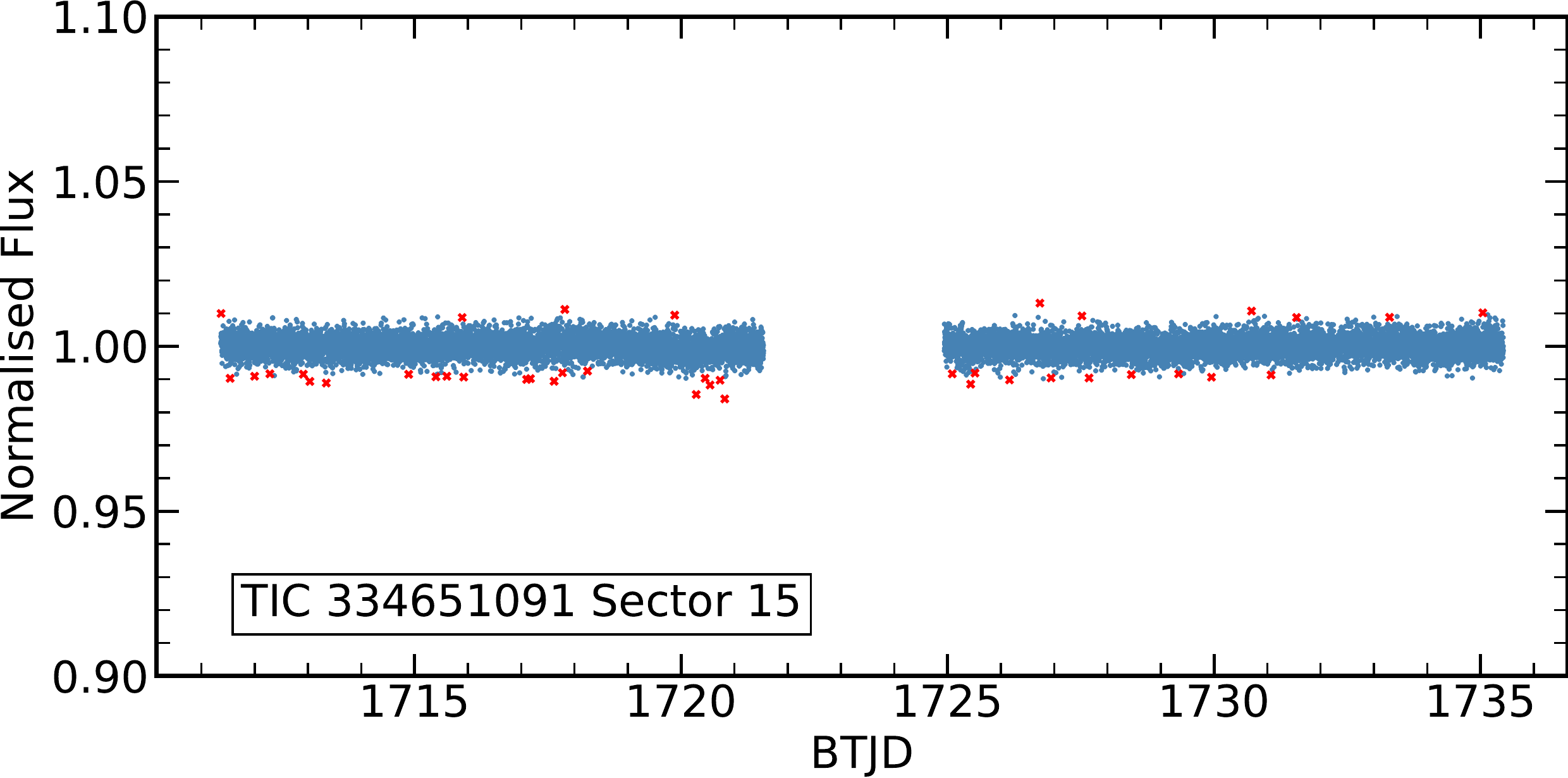}
\caption{Shown here is the SAP flux light curve of a high proper motion star G 224-79 as an example of a star with flux not variable. This is cleaned in same way as the rest of the blazars.}
\label{fig:hpms}
\end{figure}

The photometric aperture which the light curves are derived from includes extra light from nearby sources \citep{2012PASP..124..985S}. This extra flux would hinder the transit analysis for planet detection by decreasing the apparent planet transit depth. Also, there could be cases where the aperture does not include the entire Point Spread Function (PSF) of the source. For this, the SAP fluxes were modified by using the flux fraction, a parameter linked with each cadence of the light curves. This parameter gives us the fraction of the target flux contained in the optimal aperture to the total flux of the target. This along with another parameter, crowding metric, are accounted for when the light curves are processed resulting in the PDCSAP flux \citep{khb}. All the data that has been quality flagged\footnote{\url{https://outerspace.stsci.edu/display/TESS/2.0+-+Data+Product+Overview}} by \textit{TESS} team were removed. This includes data points that have been flagged for reasons such as scattered light, cosmic rays, and insufficient targets for error correction. In the next step, the outliers beyond 3 sigma in the 0.1d binned light curves for each light curve were discarded. In particular, light curves from two observations showing high and low variability are presented in the top and bottom panel, respectively, of Figure~\ref{fig:mse}. The outliers are marked in red symbols. Additional light curves are presented in Appendix~\ref{outliers-fig}. For comparison, a representative light curve of non-variable sources (star G 224-79) is also shown in Figure~\ref{fig:hpms}. 

\section{Analysis and Results}

Using the cleaned data, we have calculated fractional variability of each blazar for each sector, and optical flux distributions for each blazar have been fitted with normal and log-normal Probability Distribution Function (PDF). These plots and their corresponding fit statistics that we have calculated using Maximum Likelihood Estimator (MLE) are presented in this section. We then study the Power Spectral Density (PSD) of the blazars by modelling them using the Power Spectral Response (PSRESP) method which is a Monte Carlo technique. We use a simple power law model to compute the true power spectra of unevenly sampled light curves and compare them against the observed power spectra to produce a reliable estimate of the goodness of fit of the mentioned model as described in \citet{uttley}. The details of the methods and the corresponding results are described below.

\subsection{Fractional Variability}
\label{sec:maths} 
Blazars exhibit variability that show flux changes in timescales ranging from minutes to hours to years. In this study, we have SAP flux data for 29 blazars with two-minute cadence where the light curves are of the duration that range between 21.25 and 28.2 days. The average variability over this period can be quantified by calculating their fractional variability (FV) given by
\begin{equation}
 F_{\mathrm{var}}=\sqrt{\frac{S^2 - \langle\sigma_{err}^2\rangle}{\langle x\rangle^2}},
	\label{eq:fracvar}
\end{equation}

where $\langle\sigma_{err}^2\rangle$ is the mean square error and is given by
\begin{equation}
 \langle\sigma_{err}^2\rangle = \frac{1}{N} \sum_{i=1}^N \sigma_{err,i}^2
\end{equation} and $\sigma_{err,i}$ gives the measurement uncertainties of flux values.
\\
\\
The uncertainty in $F_{var}$ is given by
\begin{equation}
 \Delta F_{\mathrm{var}} = \sqrt{F_{\mathrm{var}}^2 + err(\sigma_\mathrm{NXS}^2)} - F_{\mathrm{var}}
\end{equation}
\\
Where err($\sigma_\mathrm{NXS}^2$) is given by
\begin{equation}
 err(\sigma_\mathrm{NXS}^2) =\sqrt {\Bigg( \sqrt{\frac{2}{N}} \frac{\langle\sigma_{err}^2\rangle}{\langle x\rangle^2}\Bigg) + \Bigg(\sqrt{\frac{\langle\sigma_{err}^2\rangle}{N}} \frac{2F_{\mathrm{var}}}{\langle x\rangle}\Bigg)^2}
\end{equation}
\cite{2003MNRAS.345.1271V}, \cite{galaxies6010002}

The FV of the sample sources are listed in 9th column of Table~\ref{tab:primary}. The result of the analysis shows that the mean fractional variability of the sources is 13.44\% with a standard deviation of 12.37\%, and that the source S4 0954+65 with over 50\% FV is found to be the most variable source whereas 3C 273 with ${0.82 \pm 2\%}$ is the least variable of the sample sources.


\begin{figure}
\centering
\includegraphics[width=\columnwidth]{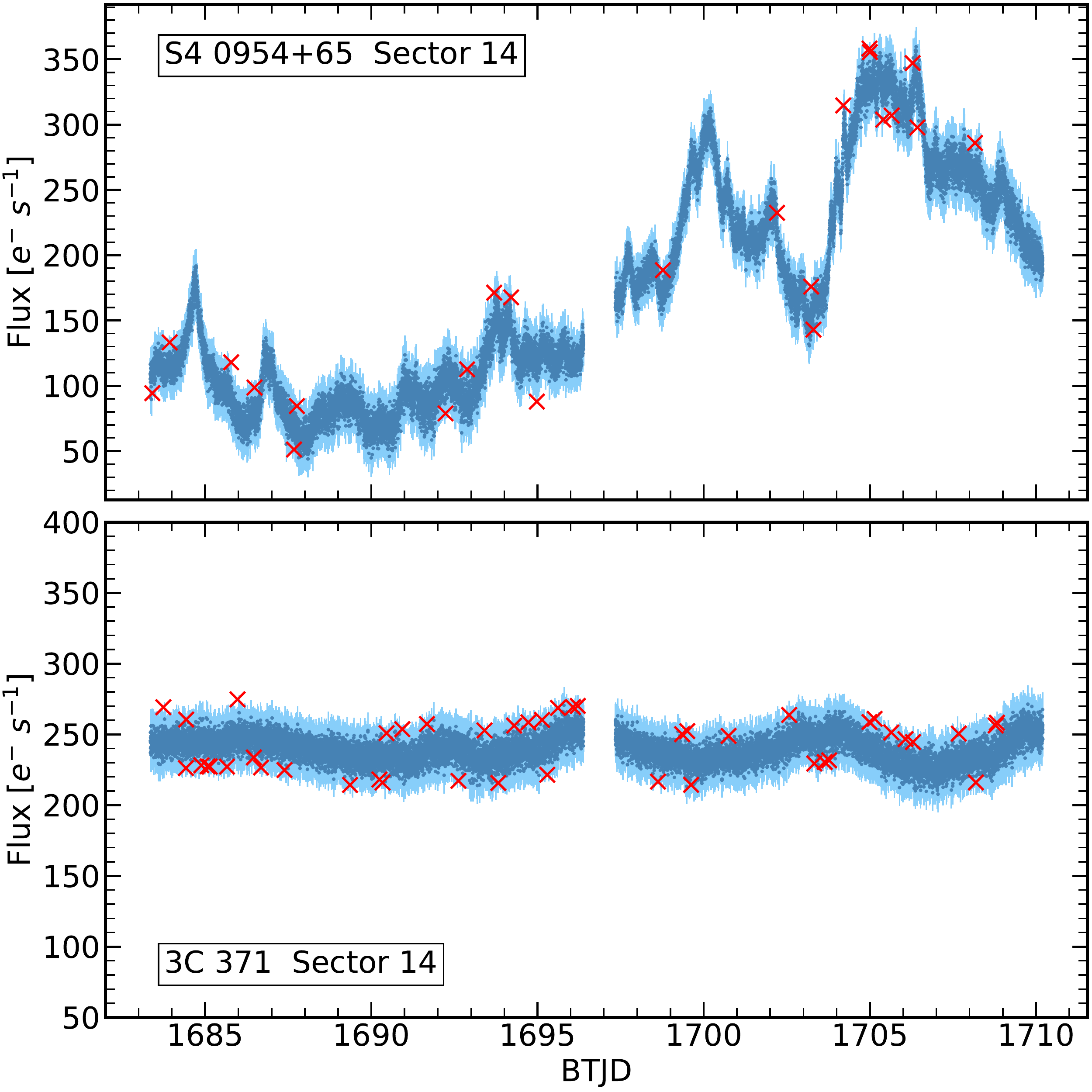}
\caption{The top-panel shows a light curve where the mean square error is lower than the variance of the flux while the bottom-panel shows where the mean square error is greater.}
\label{fig:mse}
\end{figure}


\begin{table}
\centering
 \caption{The observation periods for the sectors concerned with table~\ref{tab:primary}}
 \label{tab:sectordates}
 \begin{tabular}{llc}
Sector & Dates of Observation & Hemisphere
\\
\hline
14  &  07/18/19-08/15/19  &  N \\
15  &  08/15/19-09/11/19  &  N \\
16  &  09/11/19-10/07/19  &  N \\
17  &  10/07/19-11/02/19  &  N \\
18  &  11/02/19-11/27/19  &  N \\
19  &  11/27/19-12/24/19  &  N \\
20  &  12/24/19-01/21/20  &  N \\
21  &  01/21/20-02/18/20  &  N \\
22  &  02/18/20-03/18/20  &  N \\
23  &  03/18/20-04/16/20  &  N \\
24  &  04/16/20-05/13/20  &  N \\
25  &  05/13/20-06/08/20  &  N \\
26  &  06/08/20-07/04/20  &  N \\
27  &  07/04/20-07/30/20  &  S \\
28  &  07/30/20-08/26/20  &  S \\
29  &  08/26/20-09/22/20  &  S \\
30  &  09/22/20-10/21/20  &  S \\
31  &  10/21/20-11/19/20  &  S \\
32  &  11/19/20-12/17/20  &  S \\
33  &  12/17/20-01/13/21  &  S \\
34  &  01/13/21-02/09/21  &  S \\
35  &  02/09/21-03/07/21  &  S \\
36  &  03/07/21-04/02/21  &  S \\
37  &  04/02/21-04/28/21  &  S \\
38  &  04/28/21-05/26/21  &  S \\
39  &  05/26/21-06/24/21  &  S \\
40  &  06/24/21-07/23/21  &  N \\
41  &  07/23/21-08/20/21  &  N \\
44  &  10/12/21-11/06/21  &  N \\
45  &  11/06/21-12/02/21  &  N \\
46  &  12/02/21-12/30/21  &  N \\
\\
\hline
 \end{tabular}
\end{table}

\begin{figure}
 \centering
\includegraphics[width=\columnwidth]{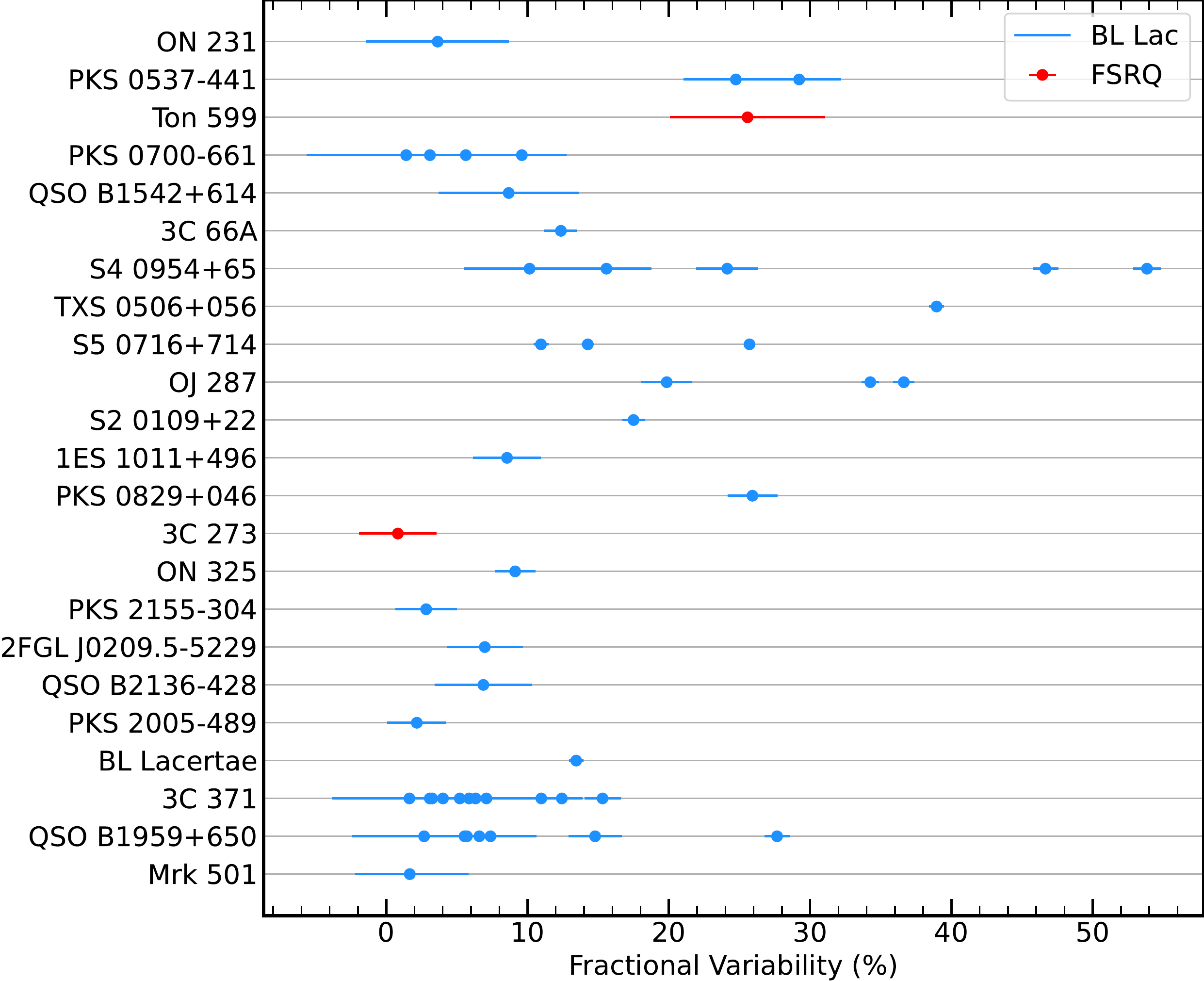}
\caption{Fractional variability of the source light curves, when mean squared error is less than the variance of flux. The sources are ordered in the vertical axis in the order of increasing corresponding red-shifts.}
\label{fig:lineplt}
\end{figure}

\onecolumn

\begin{center}
\begin{longtable}{llllclccc}
 \caption{Descriptive statistics for the flux along with spectral indices.}
 \label{tab:primary}
 \\
 \hline
 Source & R.A. & Dec. & Class & Sector & Mean ($e^{-}s^{-1}$) & S.D. ($e^{-}s^{-1}$) & FV (\%) & $\alpha$
 \\
 \hline
\multirow{ 13 } {*} { 3C 371 } & \multirow{ 13 } {*} {$ 271^{\circ}.711 $} & \multirow{ 13 } {*} {$ 69^{\circ}.8245 $} & \multirow{ 13 } {*} { BL Lac } & 14 & 239.45 & 9.08 & - & 2.0$\pm$0.23\\ 
& & & & 15 & 229.24 & 11.69 & - & 1.9$\pm$0.58\\ 
& & & & 16 & 246.77 & 16.29 & 4.01$\pm$4 & 2.0$\pm$0.05\\ 
& & & & 17 & 221.95 & 15.85 & 5.86$\pm$3 & 2.3$\pm$0.63\\ 
& & & & 18 & 256.45 & 33.16 & 12.43$\pm$1 & 3.1$\pm$0.46\\ 
& & & & 20 & 275.63 & 20.67 & 6.32$\pm$2 & 2.2$\pm$0.37\\ 
& & & & 21 & 279.33 & 44.2 & 15.31$\pm$1 & 3.0$\pm$0.55\\ 
& & & & 22 & 282.14 & 33.17 & 10.98$\pm$1 & 2.2$\pm$0.76\\ 
& & & & 23 & 270.83 & 21.43 & 7.09$\pm$2 & 2.0$\pm$1.02\\ 
& & & & 24 & 262.8 & 18.25 & 5.2$\pm$3 & 2.1$\pm$0.77\\ 
& & & & 25 & 263.43 & 14.37 & 3.08$\pm$4 & 1.9$\pm$0.59\\ 
& & & & 26 & 267.05 & 12.94 & 1.64$\pm$5 & 2.1$\pm$0.28\\ 
& & & & 41 & 273.43 & 12.54 & 3.25$\pm$3 & 2.2$\pm$0.38\\ 
\hline
\multirow{ 13 } {*} { PKS 0700-661 } & \multirow{ 13 } {*} {$ 105^{\circ}.13 $} & \multirow{ 13 } {*} {$ -66^{\circ}.1792 $} & \multirow{ 13 } {*} { BL Lac } & 27 & 98.35 & 5.42 & - & 2.3$\pm$0.28\\ 
& & & & 28 & 99.98 & 7.96 & - & 2.4$\pm$0.66\\ 
& & & & 29 & 109.77 & 6.9 & - & 2.6$\pm$0.4\\ 
& & & & 30 & 82.59 & 6.07 & - & 2.1$\pm$0.27\\ 
& & & & 31 & 87.99 & 10.29 & 9.6$\pm$3 & 2.9$\pm$0.54\\ 
& & & & 32 & 89.96 & 5.5 & - & 2.6$\pm$0.33\\ 
& & & & 33 & 106.68 & 5.38 & - & 2.0$\pm$0.21\\ 
& & & & 34 & 97.48 & 5.3 & - & 1.9$\pm$0.21\\ 
& & & & 35 & 91.36 & 5.24 & - & 1.9$\pm$0.35\\ 
& & & & 36 & 100.12 & 6.67 & 1.41$\pm$7 & 2.5$\pm$0.42\\ 
& & & & 37 & 96.59 & 8.64 & 5.63$\pm$4 & 2.0$\pm$0.21\\ 
& & & & 38 & 98.15 & 5.26 & - & 2.0$\pm$0.21\\ 
& & & & 39 & 112.28 & 8.32 & 3.09$\pm$5 & 1.8$\pm$0.47\\ 
\hline
\multirow{ 11 } {*} { QSO B1959+650 } & \multirow{ 11 } {*} {$ 299^{\circ}.999 $} & \multirow{ 11 } {*} {$ 65^{\circ}.1485 $} & \multirow{ 11 } {*} { BL Lac } & 14 & 240.56 & 22.51 & 7.38$\pm$3 & 2.0$\pm$0.83\\ 
& & & & 15 & 245.64 & 38.57 & 14.78$\pm$1 & 3.1$\pm$0.68\\ 
& & & & 16 & 224.99 & 17.07 & 5.71$\pm$3 & 2.7$\pm$0.45\\ 
& & & & 17 & 240.75 & 13.35 & 2.68$\pm$5 & 2.8$\pm$0.71\\ 
& & & & 18 & 244.67 & 22.66 & - & 2.9$\pm$0.46\\ 
& & & & 19 & 220.26 & 9.78 & - & 2.2$\pm$0.14\\ 
& & & & 21 & 246.85 & 69.32 & 27.66$\pm$0 & 2.9$\pm$0.02\\ 
& & & & 22 & 232.73 & 21.26 & 6.59$\pm$3 & 2.8$\pm$0.6\\ 
& & & & 24 & 207.48 & 12.7 & - & 2.7$\pm$0.54\\ 
& & & & 25 & 209.56 & 16.58 & - & 2.5$\pm$0.73\\ 
& & & & 41 & 263.68 & 17.74 & 5.54$\pm$2 & 2.2$\pm$0.95\\ 
\hline
\multirow{ 6 } {*} { QSO B1542+614 } & \multirow{ 6 } {*} {$ 235^{\circ}.737 $} & \multirow{ 6 } {*} {$ 61^{\circ}.4987 $} & \multirow{ 6 } {*} { BL Lac } & 15 & 97.58 & 8.87 & - & 2.5$\pm$0.61\\ 
& & & & 16 & 77.34 & 8.87 & - & 2.5$\pm$0.71\\ 
& & & & 21 & 80.95 & 7.09 & - & 3.2$\pm$0.72\\ 
& & & & 22 & 103.98 & 14.23 & 8.67$\pm$4 & 2.7$\pm$0.46\\ 
& & & & 23 & 79.54 & 9.51 & - & 2.9$\pm$0.5\\ 
& & & & 24 & 76.6 & 10.49 & - & 2.2$\pm$0.33\\ 
\hline
\multirow{ 5 } {*} { S4 0954+65 } & \multirow{ 5 } {*} {$ 149^{\circ}.697 $} & \multirow{ 5 } {*} {$ 65^{\circ}.5652 $} & \multirow{ 5 } {*} { BL Lac } & 14 & 170.79 & 80.94 & 46.65$\pm$0 & 2.9$\pm$0.58\\ 
& & & & 20 & 110.55 & 29.16 & 24.13$\pm$2 & 2.4$\pm$0.52\\ 
& & & & 21 & 110.31 & 60.49 & 53.84$\pm$0 & 2.6$\pm$0.67\\ 
& & & & 40 & 90.95 & 13.99 & 10.14$\pm$4 & 1.8$\pm$0.17\\ 
& & & & 41 & 108.97 & 20.62 & 15.59$\pm$3 & 2.1$\pm$0.95\\ 
\hline
\multirow{ 3 } {*} { OJ 287 } & \multirow{ 3 } {*} {$ 133^{\circ}.704 $} & \multirow{ 3 } {*} {$ 20^{\circ}.1085 $} & \multirow{ 3 } {*} { BL Lac } & 44 & 238.33 & 82.19 & 34.26$\pm$0 & 2.8$\pm$0.22\\ 
& & & & 45 & 212.12 & 78.51 & 36.63$\pm$0 & 2.9$\pm$0.29\\ 
& & & & 46 & 139.38 & 29.39 & 19.85$\pm$1 & 3.1$\pm$0.37\\ 
\hline
\multirow{ 3 } {*} { S5 0716+714 } & \multirow{ 3 } {*} {$ 110^{\circ}.473 $} & \multirow{ 3 } {*} {$ 71^{\circ}.3434 $} & \multirow{ 3 } {*} { BL Lac } & 20 & 2267.91 & 583.2 & 25.71$\pm$0 & 2.8$\pm$0.54\\ 
& & & & 26 & 1336.73 & 191.36 & 14.26$\pm$0 & 2.1$\pm$0.66\\ 
& & & & 40 & 979.37 & 107.86 & 10.95$\pm$0 & 1.7$\pm$0.28\\ 
\hline
\multirow{ 2 } {*} { PKS 0537-441 } & \multirow{ 2 } {*} {$ 84^{\circ}.7098 $} & \multirow{ 2 } {*} {$ -44^{\circ}.0858 $} & \multirow{ 2 } {*} { BL Lac } & 32 & 32.77 & 10.23 & 24.74$\pm$3 & 3.0$\pm$0.46\\ 
& & & & 33 & 38.15 & 13.04 & 29.22$\pm$2 & 2.2$\pm$0.95\\ 
\hline
\multirow{ 2 } {*} { Mrk 501 } & \multirow{ 2 } {*} {$ 253^{\circ}.468 $} & \multirow{ 2 } {*} {$ 39^{\circ}.7602 $} & \multirow{ 2 } {*} { BL Lac } & 24 & 496.01 & 18.58 & 2.66$\pm$3 & 2.9$\pm$0.44\\ 
& & & & 25 & 519.58 & 16.5 & 1.67$\pm$3 & 2.7$\pm$0.39\\ 
\hline
\multirow{ 2 } {*} { 2FGL J0209.5-5229 } & \multirow{ 2 } {*} {$ 32^{\circ}.3401 $} & \multirow{ 2 } {*} {$ -52^{\circ}.4897 $} & \multirow{ 2 } {*} { BL Lac } & 29 & 196.69 & 15.99 & 6.98$\pm$2 & 2.9$\pm$0.47\\ 
& & & & 30 & 192.72 & 10.77 & - & 2.3$\pm$0.59\\ 
\hline
\multirow{ 1 } {*} { Ton 599 } & \multirow{ 1 } {*} {$ 179^{\circ}.883 $} & \multirow{ 1 } {*} {$ 29^{\circ}.2455 $} & \multirow{ 1 } {*} { FSRQ } & 22 & 37.73 & 14.7 & 25.57$\pm$5 & 2.3$\pm$0.91\\ 
\hline
\multirow{ 1 } {*} { TXS 0506+056 } & \multirow{ 1 } {*} {$ 77^{\circ}.3582 $} & \multirow{ 1 } {*} {$ 5^{\circ}.69314 $} & \multirow{ 1 } {*} { BL Lac } & 32 & 277.27 & 108.57 & 38.95$\pm$0 & 2.7$\pm$0.51\\ 
\hline
\multirow{ 1 } {*} { PKS 2155-304 } & \multirow{ 1 } {*} {$ 329^{\circ}.717 $} & \multirow{ 1 } {*} {$ -30^{\circ}.2256 $} & \multirow{ 1 } {*} { BL Lac } & 28 & 786.46 & 25.32 & 2.82$\pm$2 & 2.2$\pm$0.75\\ 
\hline
\multirow{ 1 } {*} { QSO B2136-428 } & \multirow{ 1 } {*} {$ 324^{\circ}.851 $} & \multirow{ 1 } {*} {$ -42^{\circ}.589 $} & \multirow{ 1 } {*} { BL Lac } & 28 & 143.03 & 12.48 & 6.87$\pm$3 & 1.9$\pm$0.52\\ 
\hline
\multirow{ 1 } {*} { PKS 2005-489 } & \multirow{ 1 } {*} {$ 302^{\circ}.356 $} & \multirow{ 1 } {*} {$ -48^{\circ}.8316 $} & \multirow{ 1 } {*} { BL Lac } & 27 & 904.31 & 22.32 & 2.17$\pm$2 & 3.1$\pm$0.54\\ 
\hline
\multirow{ 1 } {*} { PKS 0829+046 } & \multirow{ 1 } {*} {$ 127^{\circ}.954 $} & \multirow{ 1 } {*} {$ 4^{\circ}.49418 $} & \multirow{ 1 } {*} { BL Lac } & 34 & 152.1 & 41.76 & 25.91$\pm$1 & 2.5$\pm$0.54\\ 
\hline
\multirow{ 1 } {*} { PKS 0422+004 } & \multirow{ 1 } {*} {$ 66^{\circ}.1952 $} & \multirow{ 1 } {*} {$ 0^{\circ}.601754 $} & \multirow{ 1 } {*} { BL Lac } & 32 & 99.96 & 6.43 & - & 1.9$\pm$0.19\\ 
\hline
\multirow{ 1 } {*} { 1ES 0414+009 } & \multirow{ 1 } {*} {$ 64^{\circ}.2187 $} & \multirow{ 1 } {*} {$ 1^{\circ}.08997 $} & \multirow{ 1 } {*} { BL Lac } & 32 & 71.25 & 4.82 & - & 1.9$\pm$0.2\\ 
\hline
\multirow{ 1 } {*} { PKS 0301-243 } & \multirow{ 1 } {*} {$ 45^{\circ}.8604 $} & \multirow{ 1 } {*} {$ -24^{\circ}.1198 $} & \multirow{ 1 } {*} { BL Lac } & 31 & 134.2 & 5.33 & - & 2.5$\pm$0.28\\ 
\hline
\multirow{ 1 } {*} { 1Jy0118-272 } & \multirow{ 1 } {*} {$ 20^{\circ}.1319 $} & \multirow{ 1 } {*} {$ -27^{\circ}.0235 $} & \multirow{ 1 } {*} { BL Lac } & 30 & 112.35 & 4.5 & - & 1.7$\pm$0.18\\ 
\hline
\multirow{ 1 } {*} { S2 0109+22 } & \multirow{ 1 } {*} {$ 18^{\circ}.0243 $} & \multirow{ 1 } {*} {$ 22^{\circ}.7441 $} & \multirow{ 1 } {*} { BL Lac } & 17 & 615.04 & 108.81 & 17.5$\pm$0 & 2.5$\pm$0.66\\ 
\hline
\multirow{ 1 } {*} { BL Lacertae } & \multirow{ 1 } {*} {$ 330^{\circ}.68 $} & \multirow{ 1 } {*} {$ 42^{\circ}.2778 $} & \multirow{ 1 } {*} { BL Lac } & 16 & 1169.98 & 158.04 & 13.44$\pm$0 & 2.1$\pm$0.64\\ 
\hline
\multirow{ 1 } {*} { PKS 1424+240 } & \multirow{ 1 } {*} {$ 216^{\circ}.752 $} & \multirow{ 1 } {*} {$ 23^{\circ}.8 $} & \multirow{ 1 } {*} { BL Lac } & 23 & 338.28 & 8.86 & - & 2.2$\pm$0.36\\ 
\hline
\multirow{ 1 } {*} { ON 231 } & \multirow{ 1 } {*} {$ 185^{\circ}.382 $} & \multirow{ 1 } {*} {$ 28^{\circ}.2329 $} & \multirow{ 1 } {*} { BL Lac } & 22 & 192.18 & 13.3 & 3.64$\pm$5 & 3.0$\pm$0.7\\ 
\hline
\multirow{ 1 } {*} { ON 325 } & \multirow{ 1 } {*} {$ 184^{\circ}.467 $} & \multirow{ 1 } {*} {$ 30^{\circ}.1168 $} & \multirow{ 1 } {*} { BL Lac } & 22 & 400.58 & 38.09 & 9.12$\pm$1 & 2.5$\pm$0.58\\ 
\hline
\multirow{ 1 } {*} { 1ES 1011+496 } & \multirow{ 1 } {*} {$ 153^{\circ}.767 $} & \multirow{ 1 } {*} {$ 49^{\circ}.4335 $} & \multirow{ 1 } {*} { BL Lac } & 21 & 202.78 & 19.67 & 8.55$\pm$2 & 3.0$\pm$0.57\\ 
\hline
\multirow{ 1 } {*} { PKS 1222+216 } & \multirow{ 1 } {*} {$ 186^{\circ}.227 $} & \multirow{ 1 } {*} {$ 21^{\circ}.3795 $} & \multirow{ 1 } {*} { FSRQ } & 22 & 147.24 & 7.13 & - & 2.6$\pm$0.32\\ 
\hline
\multirow{ 1 } {*} { 3C 273 } & \multirow{ 1 } {*} {$ 187^{\circ}.278 $} & \multirow{ 1 } {*} {$ 2^{\circ}.05239 $} & \multirow{ 1 } {*} { FSRQ } & 46 & 943.71 & 13.38 & 0.82$\pm$2 & 2.3$\pm$0.23\\ 
\hline
\multirow{ 1 } {*} { 3C 66A } & \multirow{ 1 } {*} {$ 35^{\circ}.6651 $} & \multirow{ 1 } {*} {$ 43^{\circ}.0355 $} & \multirow{ 1 } {*} { BL Lac } & 18 & 400.4 & 50.75 & 12.36$\pm$1 & 2.6$\pm$0.65\\ 
\hline

\end{longtable}
\end{center}

\begin{multicols}{2}

\subsection{Flux Distribution: Lognormality and Normality}
Flux variability in AGNs across the electromagnetic spectrum is frequently observed in their MWL observations; however the details that describe the physical processes leading to such variability are still debated. Study of the flux distribution characterized by suitable probability density function (PDF) is one of the most important tools to constraint the physical processes. Blazar MWL, e. g. optical, X-ray and gamma-ray, flux distribution are mostly represented by normal and lognormal processes \citep[see e. g.][]{2022MNRAS.510.5280M,2021ApJ...923....7B,bhatta2020nature}. A normal flux distribution might indicate a linear summation of components contributing to the total observed emission. On the other hand, lognormal distributions i.e. distributions which are Gaussian in the logarithm of the flux, can be obtained via multiplicative (or cascade) like processes \citep{uttley2005non}. Here we construct source flux histograms using TESS observations and subsequently the histograms are fitted with normal and lognormal PDFs.

A normal PDF is given by
\begin{equation}
 f_\text{normal}(x) = \frac{1}{\sigma\sqrt{2\pi}}\mathrm{exp}\bigg( - \frac{(x-\mu)^2}{2\sigma^2}\bigg)
\end{equation} where $\mu$ and $\sigma$ are the mean and the standard deviation of the normal distribution, respectively.

Similarly the lognormal PDF is written as
\begin{equation}
 f_\text{lognormal}(x) = \frac{1}{xs\sqrt{2\pi}} \mathrm{exp} \bigg( - \frac{(\ln x-m)^2}{2 s^2} \bigg)
\end{equation} where m and s are the mean location (expressed in units of natural log) and the scale parameters of the distribution, respectively.

We modeled the blazar optical flux distributions of the sample sources using a linear combination of the two PDFs. The method adopted here employs the expectation-maximization algorithm, which is a method for performing maximum likelihood estimation in the presence of latent variables \citep{sklearn_api}. For fitting lognormal models, we have transformed the data with natural logarithm and applied normal models. The Akaike's Information Criteria (AIC) and Bayesian Information Criteria (BIC) were used to compare both normal and the lognormal models. These estimators are given as
\begin{equation}
 \mathrm{AIC} = 2 \textit{k} - 2\ln(L)
\end{equation}
\vspace{-2em}
\begin{equation}
 \mathrm{BIC} = \textit{k} \ln(n) - 2\ln(L),
\end{equation} where k is the number of estimated parameters in the model, L is the maximum value of the likelihood function, and n is the number of observations or the sample size.
AIC is an estimator that predicts error in model selection and hence is a relative quantity to determine best fit models on a given data set. BIC is a criteria for model selection from a finite group of models. Models with lower AIC and BIC values better estimate a given data than other models. The AIC and BIC values we calculated by applying normal model to log-transformed data were missing the Jacobian component of log transformation. Hence, before comparing the AIC and BIC values of both the models, we have added $2\Sigma \ln({flux})$ to the AIC and BIC of the lognormal model \citep{AIC}.

For 27 light curves of 13 blazars the AIC and BIC agreed on the same model for the flux distribution, which is either a single component or a two-component model. The bi-modal Gaussian distribution was the most prevalent model among the flux distributions of these light curves, accounting for 17 of them. The rest of the flux distributions were comprised of single component Gaussian, single component lognormal, and bi-modal lognormal models in similar proportion. The flux histograms and their normal and lognormal fitting for selected observations showing multiple components and the corresponding residuals are shown in Figure~\ref{fig:hist1}. It is interesting to note that in the case of sources PKS 0700-661 and QSO B1542+614 with multiple light curves, the best-fit PDFs tends to vary both in terms of the components and the PDF models.


\begin{table*}
	\centering
 \caption{The best of lognormal and best of normal distribution fit statistics for the optical flux distribution using MLE method, for selected light curves. The lower values of AIC and BIC corresponding to the best-fit are highlighted.}
	\label{tab:flux-distr}
	\begin{tabular}{l|c|ccc|ccc} 
		\hline
 & & \multicolumn{2}{c|}{Normal fit} & \multicolumn{2}{c}{Lognormal fit} \\
Source & Sector & $AIC$ & $BIC$ & $AIC$ & $BIC$ \\
\hline
\multirow{ 2 } {*} { 3C 371 } & 14 & \textbf{ 133535.8 } & \textbf{ 133574.9 } & 133564.1 & 133579.8\\ 
 & 26 & 132858.3 & 132897.0 & \textbf{ 132815.0 } & \textbf{ 132853.7 }\\ 
\hline
\multirow{ 11 } {*} { PKS 0700-661 } & 27 & \textbf{ 78523.6 } & \textbf{ 78538.5 } & 78538.7 & 78575.9\\ 
 & 28 & \textbf{ 85013.1 } & \textbf{ 85050.1 } & 85029.3 & 85066.3\\ 
 & 29 & \textbf{ 89599.9 } & \textbf{ 89637.4 } & 89610.5 & 89648.0\\ 
 & 30 & \textbf{ 99463.0 } & \textbf{ 99501.3 } & 99481.9 & 99520.2\\ 
 & 32 & \textbf{ 106944.1 } & \textbf{ 106982.9 } & 106963.2 & 107001.9\\ 
 & 33 & 108256.5 & 108272.1 & \textbf{ 108223.2 } & \textbf{ 108238.7 }\\ 
 & 35 & 83553.0 & 83590.6 & \textbf{ 83540.0 } & \textbf{ 83555.0 }\\ 
 & 36 & \textbf{ 100904.3 } & \textbf{ 100942.5 } & 100921.0 & 100959.1\\ 
 & 37 & \textbf{ 102950.7 } & \textbf{ 102988.8 } & 102972.2 & 103010.2\\ 
 & 38 & 110756.4 & 110772.0 & \textbf{ 110677.7 } & \textbf{ 110693.3 }\\ 
 & 39 & \textbf{ 121667.2 } & \textbf{ 121706.0 } & 121691.7 & 121730.5\\ 
\hline
\multirow{ 4 } {*} { QSO B1542+614 } & 16 & \textbf{ 90672.3 } & \textbf{ 90709.6 } & 90804.2 & 90841.6\\ 
 & 21 & 123096.4 & 123135.5 & \textbf{ 123058.5 } & \textbf{ 123097.6 }\\ 
 & 23 & \textbf{ 92179.2 } & \textbf{ 92216.5 } & 92240.4 & 92277.6\\ 
 & 24 & 122202.5 & 122241.0 & \textbf{ 122187.0 } & \textbf{ 122225.5 }\\ 
\hline
\multirow{ 1 } {*} { PKS 0537-441 } & 32 & \textbf{ 129377.6 } & \textbf{ 129416.4 } & 129944.7 & 129983.6\\ 
\hline
\multirow{ 1 } {*} { Mrk 501 } & 25 & \textbf{ 144791.5 } & \textbf{ 144830.3 } & 144795.0 & 144833.8\\ 
\hline
\multirow{ 1 } {*} { 2FGL J0209.5-5229 } & 30 & \textbf{ 126087.6 } & \textbf{ 126126.2 } & 126105.3 & 126143.9\\ 
\hline
\multirow{ 1 } {*} { PKS 0422+004 } & 32 & \textbf{ 112436.2 } & \textbf{ 112451.7 } & 112524.7 & 112563.4\\ 
\hline
\multirow{ 1 } {*} { 1ES 0414+009 } & 32 & \textbf{ 102447.5 } & \textbf{ 102463.0 } & 102545.2 & 102560.7\\ 
\hline
\multirow{ 1 } {*} { PKS 0301-243 } & 31 & \textbf{ 102491.4 } & \textbf{ 102530.0 } & 102495.7 & 102534.3\\ 
\hline
\multirow{ 1 } {*} { 1Jy0118-272 } & 30 & \textbf{ 97048.0 } & \textbf{ 97063.4 } & 97089.5 & 97105.0\\ 
\hline
\multirow{ 1 } {*} { PKS 1424+240 } & 23 & \textbf{ 95963.9 } & \textbf{ 96001.4 } & 95969.4 & 96006.9\\ 
\hline
\multirow{ 1 } {*} { PKS 1222+216 } & 22 & \textbf{ 108918.0 } & \textbf{ 108956.4 } & 108920.9 & 108959.4\\ 
\hline
\multirow{ 1 } {*} { 3C 273 } & 46 & \textbf{ 130791.1 } & \textbf{ 130829.6 } & 130794.9 & 130833.4\\ 
\hline
	\end{tabular}
\end{table*}


\begin{figure*}
 \centering
 \includegraphics[width=\textwidth]{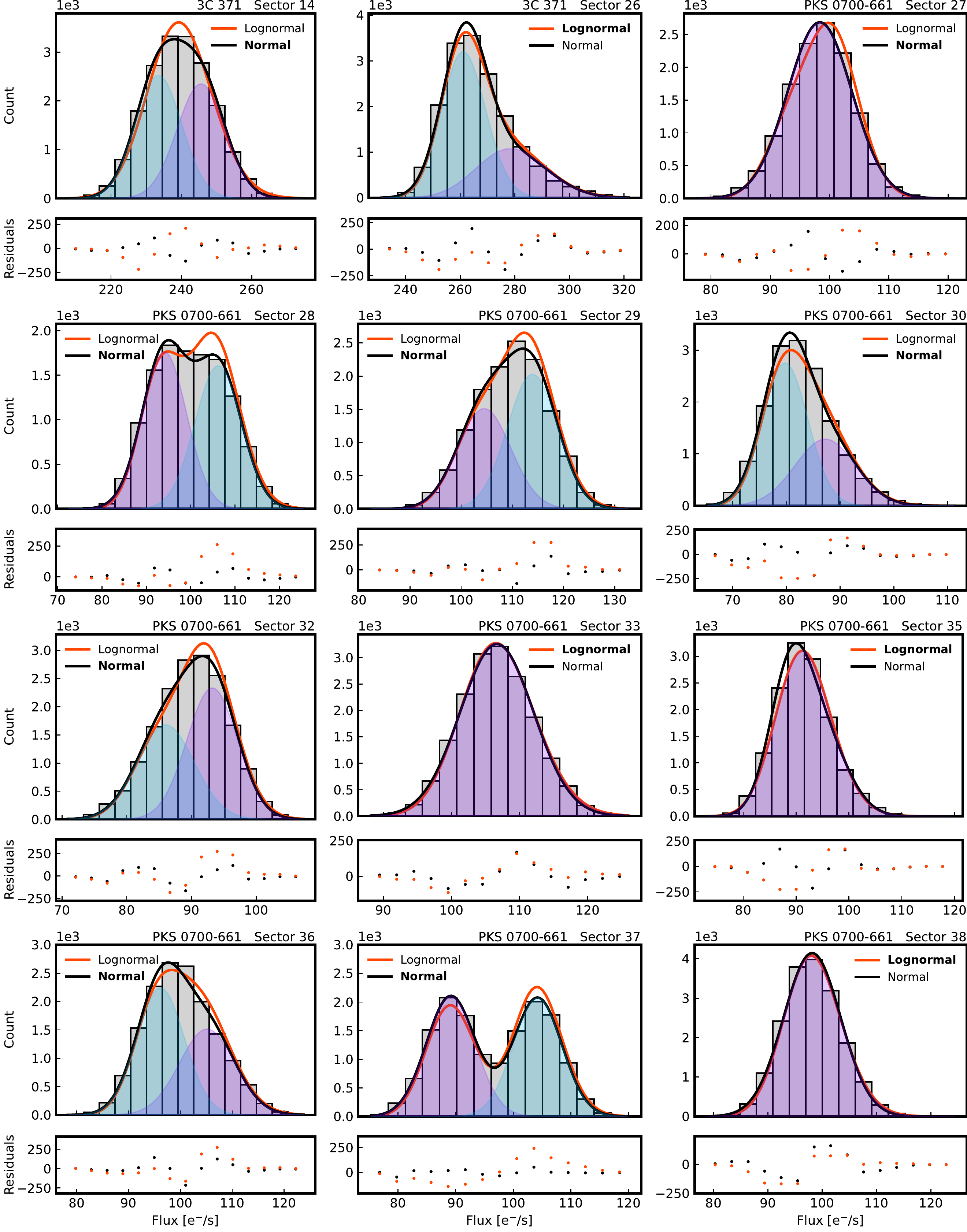}
 \caption{Normal and lognormal PDF fitted on flux histogram for the selected light curves. The colored distributions represent individual components of the best-fit model (highlighted in the legend).}
 \label{fig:hist1}
\end{figure*}

\begin{figure*}
 \centering
 \includegraphics[width=\textwidth]{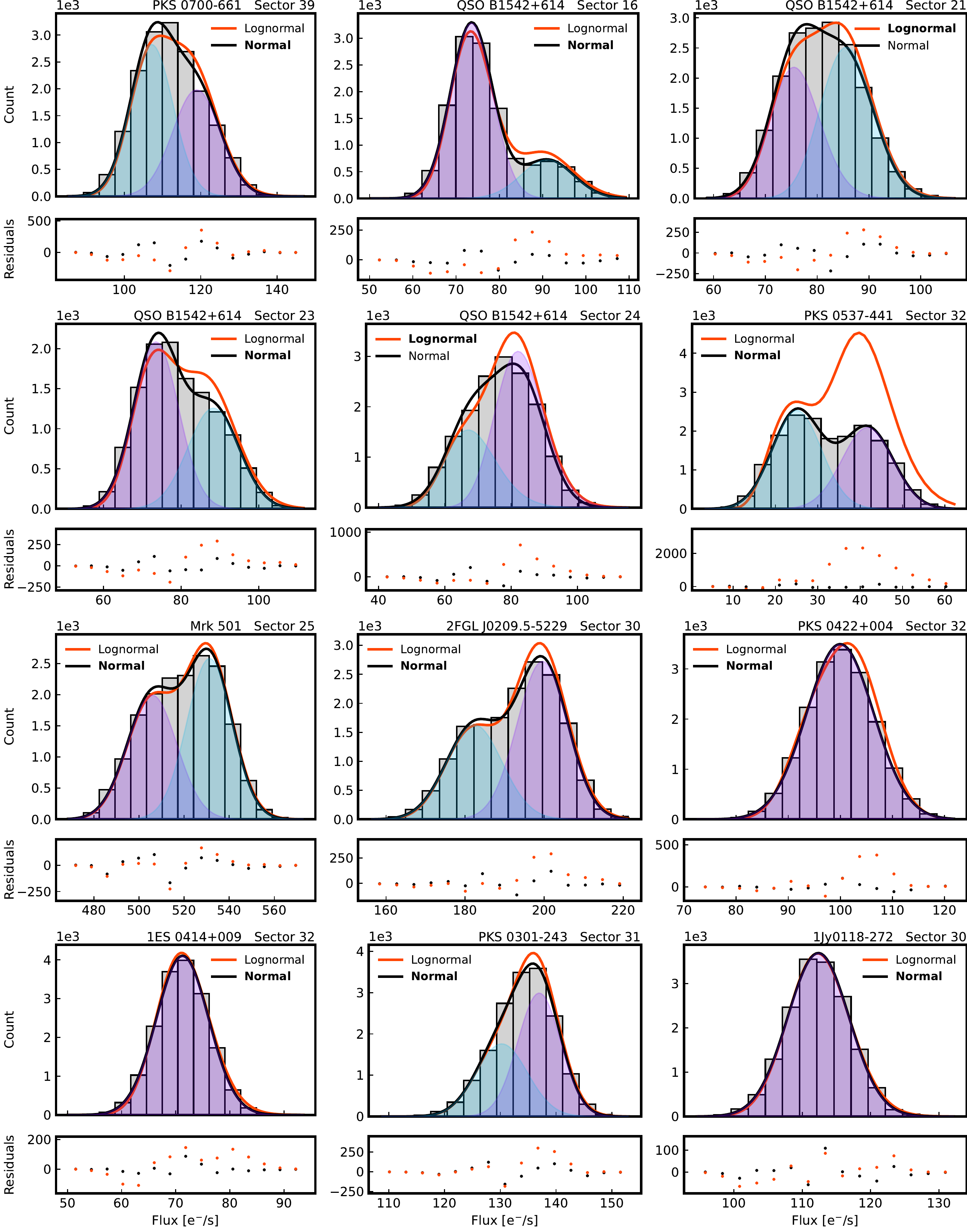}
 \contcaption{Normal and lognormal PDF fitted on flux histogram for the selected light curves. The colored distributions represent individual components of the best-fit model (highlighted in the legend).}
 \label{fig:hist2}
\end{figure*}

\begin{figure*}
 \centering
 \includegraphics[width=\textwidth]{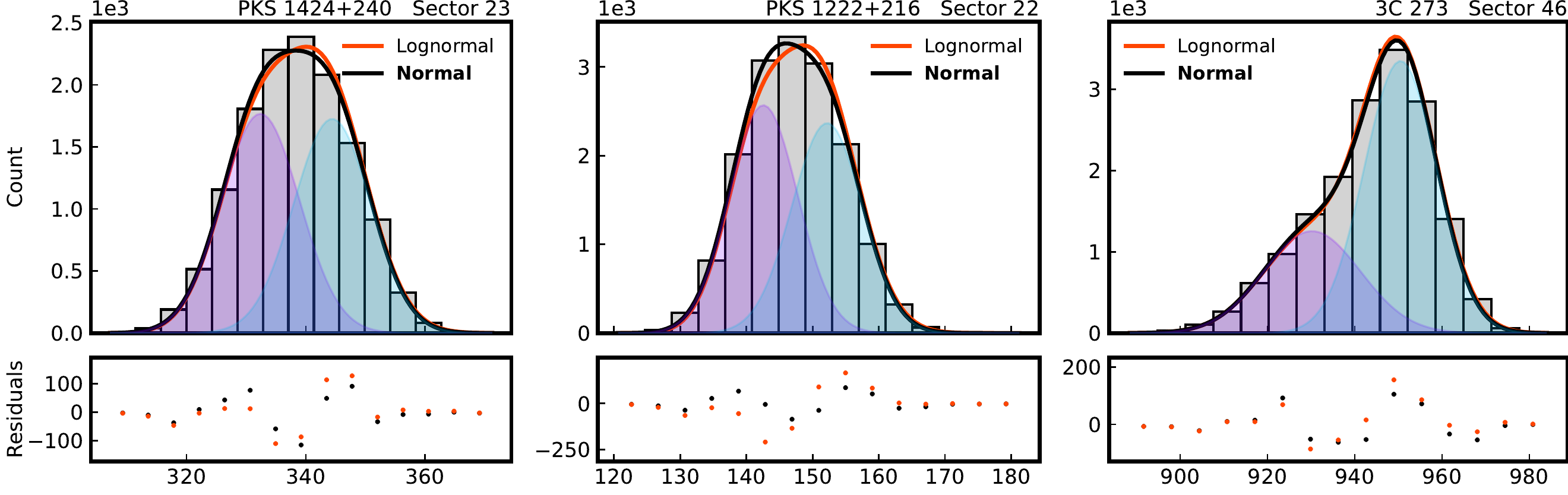}
 \contcaption{Normal and lognormal PDF fitted on flux histogram for selected light curves. The colored distributions represent individual components of the best-fit model (highlighted in the legend).}
 \label{fig:hist3}
\end{figure*}


\end{multicols}
\twocolumn

\subsection{Power Spectral Analysis}

The source light curves exhibit rapid flux variability. It is then natural to determine PSD as a function of temporal frequency, which is a measure of variability power over various timescales. The analysis is motivated to search for any possible peaks or breaks in the source PSD, which are indications of presence of timescales that could be characteristic to the system.

In general, discrete Fourier transforms can be employed to calculate PSD of the time series that are regularly sampled and that has large ratio of length of observation to sampling time interval.
The \textit{TESS} light curves are discretely sampled every 2 minutes. In some cases, the sampling times are also unevenly spaced, with a significant gap of up to 24 hours in the middle of a every single sector observation. This large gap in the middle of every \textit{TESS} light curve is caused by the observational gap that occurs when the spacecraft transmits data to the ground station on Earth during which the spacecraft is reoriented to join its Ka-band transmitter at the Earth \citep{tessinshb}. 
Aliasing and red-noise leak will impede the power density spectrum produced by such a light curve such that the periodogram can be seen to flatten out at higher frequencies \citep{uttley}. The gaps in the \textit{TESS} data make the sampling unevenly spaced such that the PSD estimation using discrete Fourier transforms would not be accurate for the given sources. In such a context, the Lomb-Scargle (LS) method \citep{lomb1976,scargle1982} proves to be a better estimation of the source PSD. Furthermore, in the case of the time-series with uneven sampling times, the Nyquist frequency is not clearly defined. Usually, taking the mean of the sampling intervals is one of the ways of calculating it \citep[see e. g.][]{scargle1982,horne,10.5555/1403886}. However, it is important to note that the frequency calculated in this manner is less than the true Nyquist frequency, beyond which all signals are aliased into the Nyquist range \citep{vanderplas}. So to construct the source PSD, the frequency components ranging from $\nu_{min}=1/{T}$ and $\nu_{max}={N}/{2T}$ were chosen, where N is the total number of data points in a light curve while T is the time baseline.

The power spectral response (PSRESP) method \citep{uttley}, that is known to remove most of the artifacts of sampling patterns, was employed to derive the intrinsic PSD of a light curve along with its associated uncertainties. To implement the method, the algorithm described in \citet{2008ApJ...689...79C} was followed. The Monte Carlo algorithm was considered to simulate the artificial light curves following the phase and amplitude randomization method described in \citet*{timmer}. However, the light curves thus generated strictly follow the Gaussian distribution; whereas the blazar flux can have other forms of distributions. Therefore, to simulate light curves of given model of PSD and PDF, the method described in \citet{Emmanoulopoulos2013}\footnote{The method was implemented using the python program based on the code at \url{https://github.com/samconnolly/DELightcurveSimulation}} was adopted. In this process, for each of the blazar light curves, 100 light curves were simulated using the best-fit PDF model from Section 4.2 and a single power-law model with the spectral indices ranging from 1.0 to 4.0 with a regular interval of 0.1. The simulated light curves were sampled at equal time intervals, and subsequently, the light curves were interpolated to mimic the \textit{TESS} light curves of each of the sources.

\begin{figure}
 \centering
 \includegraphics[width=\columnwidth]{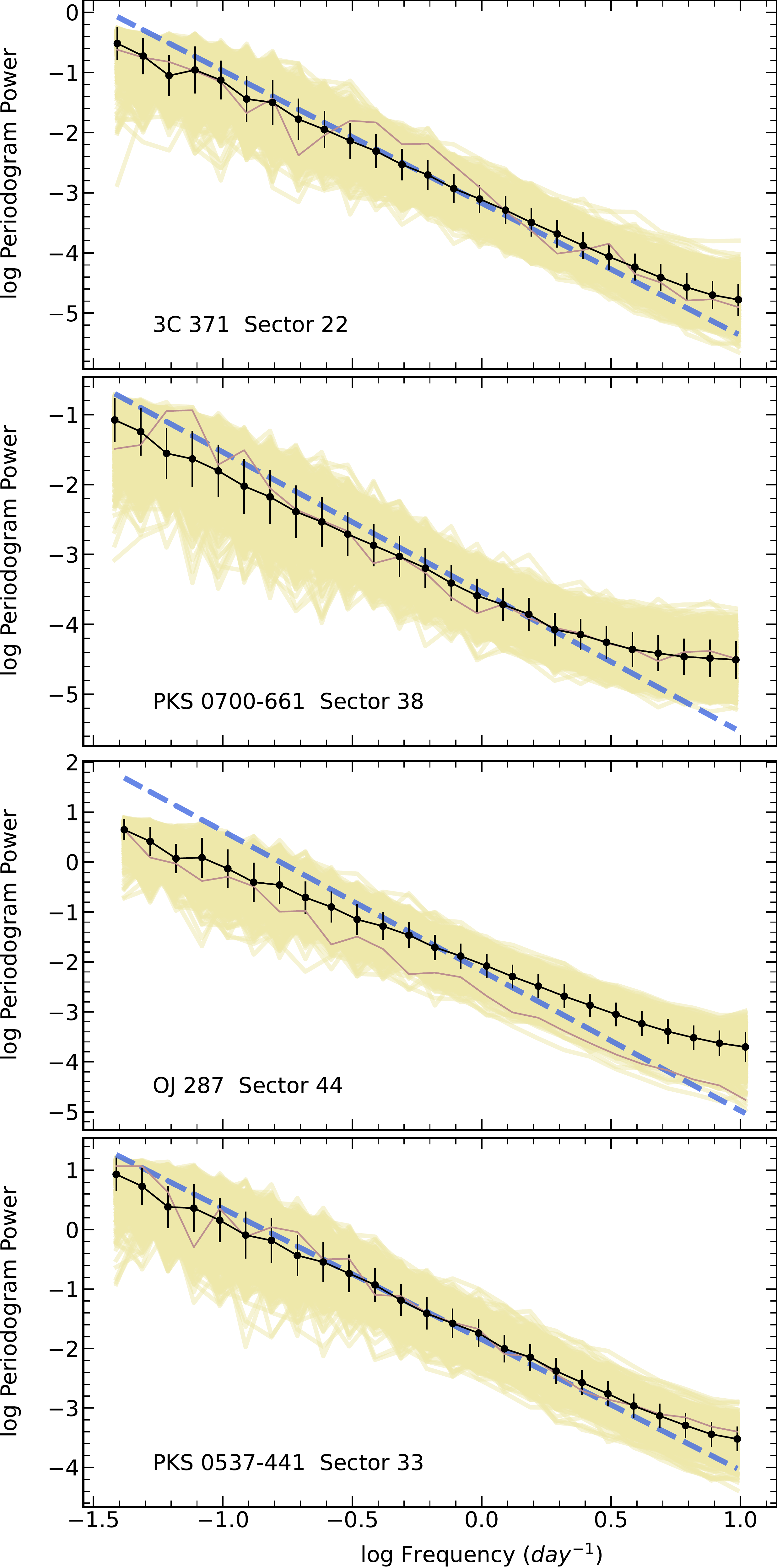}
 \caption{The best-fit power-law models along with corresponding simulated light curve’s PSD shown in relation with the observed PSD of few selected light curves. Black dotted line (with error bars) denotes the average PSD of the simulated light curves at the chosen power spectral index. The blue dashed line represents the power law model. The pink line is the observed PSD. The yellow background is the enclosed space of 100 simulations at the chosen spectral index.}
 \label{fig:psresp1}
\end{figure}

Before computing LS periodogram, the light curves were binned using 10 minute time bin and then the periodogram is computed to get the powers at frequencies up to Nyquist frequency. This is done for all the simulated light curves and a mean power spectra are calculated for all of the simulated light curves corresponding with a particular power-law model. A similar power spectrum is constructed after binning the \textit{TESS} light curve as well. Now, to determine the PSD associated with the \textit{TESS} light curve, a chi-square like statistics for the observed source periodogram was constructed as given below.
\begin{equation}
 \chi^{2}_{\mathrm{obs}} = \sum_{\nu = \nu_{\mathrm{min}}}^{\nu_\mathrm{max}} \dfrac{(PSD_{\mathrm{obs}} - \overline{PSD}_{\mathrm{sim}})^{2}} {(\Delta
PSD_{\mathrm{sim}})^{2}}
\end{equation}
A similar statistics for each of the simulated light curves was also defined as below.
\begin{equation}
 \chi^{2}_{\mathrm{dist},\mathrm{i}} = \sum_{\nu = \nu_{\mathrm{min}}}^{\nu_{\mathrm{max}}} \dfrac{(PSD_{\mathrm{sim},\mathrm{i}} - \overline{PSD}_{\mathrm{sim}})^{2}} {(\Delta PSD_{\mathrm{sim}})^{2}} 
\end{equation}
Then, the success fractions based on the magnitudes of $\chi^{2}$, are determined for each of the power-law indices by taking the ratio of total number of $\chi^{2}_{\mathrm{obs}}$ values that are smaller than $\chi^{2}_{\mathrm{dist},\mathrm{i}}$ and the total number of simulations for a simple power-law model. Then the intrinsic power spectrum of the \textit{TESS} light curve is represented by the power-law model with the highest success fraction. The LS periodogram and the best fit PSD models for a few sources are shown in Figure~\ref{fig:psresp1}.

\section{Discussion}
We performed a first systematic variability study of blazars observed by the \textit{TESS} spacecraft. As mentioned earlier, there have been several studies that monitor variability of blazars in either intra-day or sparsely sampled observation aimed to study variability properties in longer (a few years) timescales. Also, there have been some efforts to obtain continuous monitor of a single source via globally coordinated observation campaigns \citep[e.g.][]{Acciari2020,Bhatta2016,Bhatta2013}. However, such studies require considerable effort in co-coordinating campaigns demanding participation of a large number of observatories. In such context, the short cadence light curves spanning timescales of a few days provide a unique opportunity to explore variability studies in the timescales of a few days, given that most of the studies are either in intra-day timescales spanning a few hours or are of sparsely sampled long-term observations. 

In the BL Lac sources, the optical emission as detected by the instrument mostly represents non-thermal synchrotron radiation emitted by the relativistic particle moving in the jet's magnetic field. In the leptonic emission models of blazars, the energetic particles, either electrons or positrons, could have been accelerated by the shocks propagating in the large scale jets, or by magnetic reconnection events taking place in the highly magnetized region of the jets \citep{giannios2009fast}. In the internal shock scenario of the particle acceleration, the particles could be accelerated up to the energies equivalent to thousands of Lorentz factors \citep[e.g.][]{Joshi2011,Spada2001,Kirk1998,Marscher1985}.
Also, the particles might be accelerated in the high turbulent regions of the jet \citep[e.g.][]{marscher2013turbulent,calafut2015modeling}. On the other hand, in hadronic models of blazars the synchrotron emission might have resulted from the secondary particles from the hadronic interactions, such as interaction of highly accelerated protons with the ambient photon or particle fields \citep[e.g.][]{Mucke2003,Mannheim1993}. Besides the jet origin, a part of optical emission might have its origin due to the thermal emission from the accretion disk, especially in FSRQs.

The observed optical variability might have been shaped by a combination of both the source intrinsic and extrinsic variable factors. In the source intrinsic cases, variability might arise due to magnetohydrodynamical instability in the jets or the disk, e. g., due to variable injection rates at the shock front \citep{Webb2021,Xu2019,Bhatta2013}, or due to intrinsic turbulent nature of the jet \citep{Bottcher2019,Larionov2013}. In the extrinsic scenario, the observed variable flux could be a result of the geometric effects such as change in orientation, speed of the emission regions, and light crossing timescales \citep[see e.g.][]{Bhatta2018,Raiteri2017,raiteri2013awakening}.

The results from the flux distribution analysis imply that majority of flux histograms are consistent with the bi-modal normal PDF. This could be a clear evidence that the optical flux variability on timescales of a few days are mainly contributed by two different emission zones e. g. disk and jet or two dominant turbulent regions in the jet. But it also can be speculated that the days-timescale flux modulations are driven by two distinct states of the activity in a localized region of the jet. Recent results reported by \citet{2022MNRAS.510.5280M} also clearly show bi-modal normal flux distributions in the intra-day X-ray observations of blazars Mrk 501, OJ 287 and RBS 2070.
However, it should be noted that in a similar study using long-term optical observations \cite{2021ApJ...923....7B} found that the flux histograms are consistent with mainly uni-modal normal or log-normal PDFs. It appears that in the longer timescales the observed bi-modality gets smeared by the variable emission contributed by other components. As there are not many works that study the flux distribution in short timescales, the results from the current study utilizing largely uninterrupted observations from the space telescope are important in constraining the physics behind the short-term optical variability observed in blazars.


Power spectral density can reveal the stochastic nature of the variability in the form of combined emission from different realizations of a single/multiple noise-like processes. Under such a framework, the variability power spectrum can be characterized by simple power-law PSD shapes. The slope indices of the PSDs then estimate the variability amplitude over temporal frequencies. One of the primary goals of the study was to analyze the PSDs in search of the PSD breaks and potential QPOs with characteristic timescales. Such timescales corresponding to the observed break frequency can be linked to the physical parameters of the AGN system, e.g. size of the emission region, cooling timescales, etc., that contribute to the observed flux variability. Similarly, QPOs in blazar could be conceived in several scenarios \citep[see][for detailed discussion]{Bhatta2019}. However, we did not detect any significant PSD breaks in the \textit{TESS} blazar observations.

Furthermore, a number of studies involving multi-wavelength observations estimate the PSD and the corresponding slope index using mostly simple power-law of the form of $P(\nu)\propto \nu^{-\beta}$: A study including long-term light curves of 31 blazars \cite {Nilsson2018} estimated the average PSD slope of $\sim$1.42 that is uncorrelated with the synchrotron peak frequency of the sources. \citet{Max-Moerbeck2014} in a study of the cross-correlation between the radio and the gamma-ray emission from blazar analyzed 15 GHz radio light curves of a large sample of source and found the distribution of the PSD slope (spectral indices) centered around 2.3. In Gamma-ray studies using Fermi/LAT observations, the PSD slopes are found to be nearly 1.0 \citep{bhatta2020nature,Meyer2019,Sobolewska2014}. On the other hand, some works explore the jet emission models treated with computational methods. In a simplistic hydrodynamic simulation of two-dimensional relativistic jets considering both intrinsic and extrinsic scenario, \cite{Pollack_2016} observed that the PSD slopes for the bulk velocity produced variations was found to be steeper, $\sim$2.5, compared to the turbulence induced variations at $\sim$2.0. Furthermore, the study also indicated that when both intrinsic and extrinsic scenarios were considered together, the shorter timescale variations were found to be dominated by the turbulent fluctuations contributions and the longer periods by the bulk flow changes. However, the result of \citet{calafut2015modeling} shows that the PSD slopes which are highly sensitive to the turbulent velocity are largely unaffected by the change in the viewing angle and bulk velocity. Similarly, in a more realistic case, \citet{Riordan2017} based their model on turbulence in relativistic jets that are launched by magnetically arrested accretion flows (MAD) around a rapidly rotating supermassive black hole. The PSD of the modeled synchrotron and synchrotron self-Compton variable emission is in confirmation with the observed power-law spectrum at short-term timescale. In the similar MHD simulation of large scale jets, based on MAD scenario, \citet{McKinney2012} found signatures of QPOs and PSD breaks in their simulations.

\section{Conclusions}

This work represents one of the first systematic studies of optical variability of blazars using observations that are nearly continuous in the timescales of a few days. As a quantitative measure of the variable emission from the sources, we computed fractional variability of the sample sources, which spans a large range in percentage. For example, S4 0954+65 shows variability of ${53.84 \pm 0\%}$. The lowest variability percentage is seen is 3C 273 which is ${0.82 \pm 2\%}$. The average fractional variability is 13.44\% with an SD of 12.37\%. We investigated the nature of flux distribution of the optical flux of the sources by fitting the flux histogram with normal and lognormal PDFs, and performing statistical test. The lower AIC and BIC values for the normal PDF suggest that compared to the lognormal PDF, the normal PDF better represents the flux variability of the blazars. Furthermore, some of the sources reveal bi-modal normal flux distribution, an indication of two dominant emission regions or states. We performed power spectral density analysis and estimated the power spectral slopes of the light curves. It is found that the PSD is consistent with simple power-law model and the distribution of the slope indices is centered around 2.4.

\section*{Acknowledgments}
We thank the anonymous referee for a careful and thorough review of this paper, which helped improve the quality of the work. 
GB acknowledges the financial support by Narodowe Centrum Nauki (NCN) grant UMO-2017/26/D/ST9/01178. This research made use of Lightkurve, a Python package for \textit{Kepler} and \textit{TESS} data analysis (\citet{2018ascl.soft12013L}).
 Funding for the \textit{TESS} mission is provided by NASA's Science Mission directorate. This research made use of Astropy,\footnote{http://www.astropy.org} a community-developed core Python package for Astronomy \citep{astropy:2013, astropy:2018}. This research also made use of astroquery, an astronomical web-querying package in Python \citep{2019AJ....157...98G}.

\section*{Data Availability}
The paper includes data collected by the \textit{TESS} mission, which are publicly available from the Mikulski Archive for Space Telescopes (MAST). Furthermore, the processed data will be shared on reasonable request to the corresponding author.



\bibliographystyle{mnras}
\bibliography{ref.bib} 




\appendix

\section{Light Curves}

\label{outliers-fig}

This section provides some of the light curves of blazars shown in Table~\ref{tab:primary}. These light curves contain SAP flux data that hasn't been quality flagged. The red crosses represent the outliers ignored during the analysis, based on 3 Sigma of 0.1 day bins.

\onecolumn
\begin{figure}
 \centering
\includegraphics[width=\columnwidth]{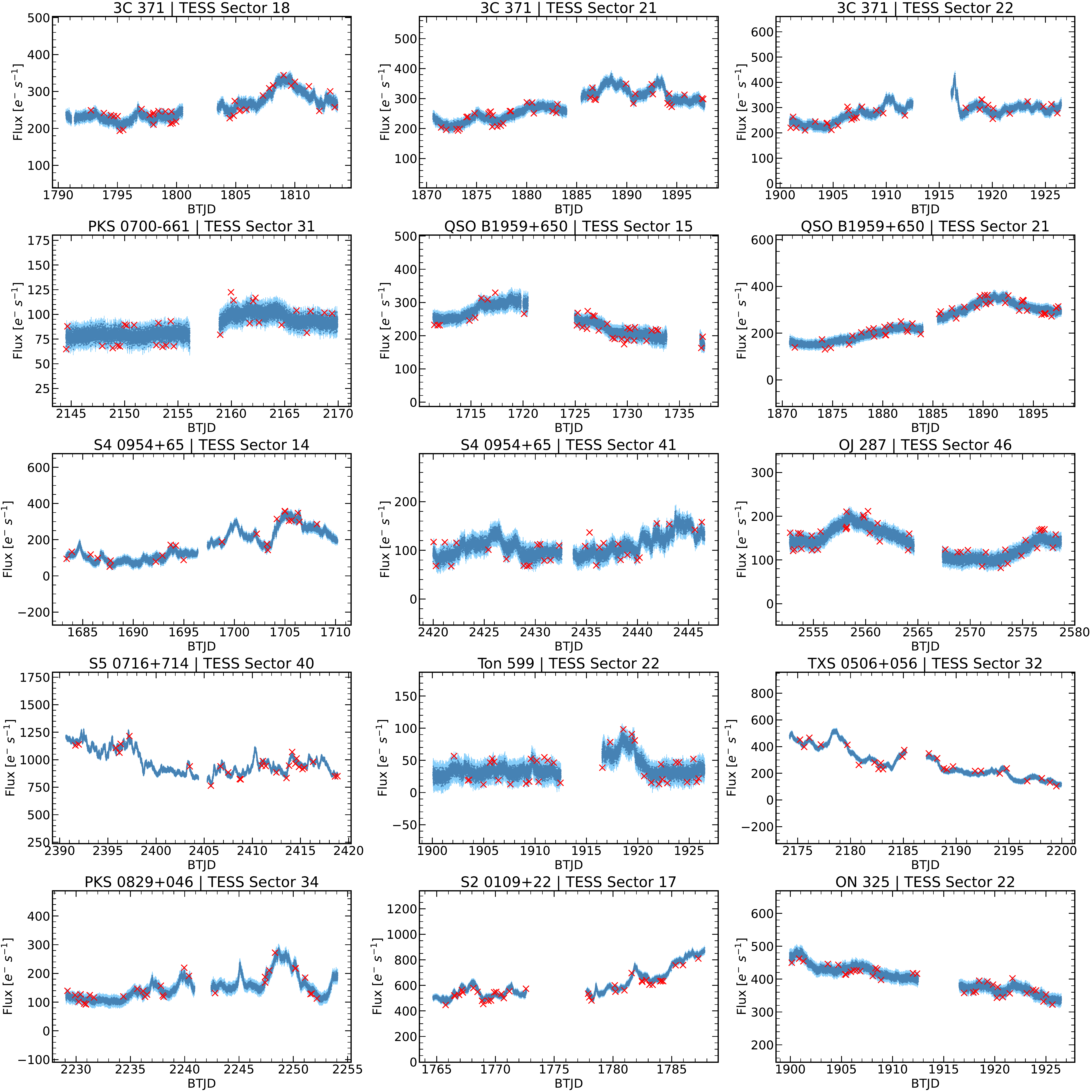}
\caption{Some light curves from Table~\ref{tab:primary}. The blue lines indicate the data that hasn't been flagged for quality issues while the red crosses indicate the outliers.}
\end{figure}


\bsp	
\label{lastpage}
\end{document}